\documentclass[12pt]{article}
\usepackage[polish,english]{babel}
\usepackage[latin1]{inputenc}
\usepackage{times}
\usepackage[T1]{fontenc}
\usepackage{epsfig}

\begin{document}

\title{Compact Shock Impulses in Models with V-shaped Potentials}

\author{P. Klimas $\;$\\$\;\;$ \\  Departamento de Fisica de Particulas,\\ Facultad de Fisica\\ Universidad de Santiago\\ E-15706 Santiago de Compostela, SPAIN}

\date{$\;$}

\maketitle

\begin{abstract}
A new class of solutions in the signum-Klein-Gordon model is presented. Our solutions merge properties of shock waves and compactons that appear in scalar field models with V-shaped potentials. 
\end{abstract}

\vspace*{2cm} \noindent PACS: 03.50.Kk, 05.45.-a, 11.10.Lm \\
\

\pagebreak

\section{Introduction}
Scalar field-theoretic models play an important role in contemporary theoretical physics. They have a wide range of applications from condensed matter physics \cite{condensed}  to cosmology \cite{cosmology}. An interesting new class of scalar field models has been proposed in   
\cite{AKTActa36}.  Such models are called models with V-shaped potentials due to their common property of  a non-vanishing first derivative of the potential at the minimum. This feature introduces a qualitatively new behaviour of fields close to the vacuum. It turns out that for models with V-shaped potentials the field approaches its vacuum value in a polynomial (quadratic) way.  As a consequence, topological defects ({\it e.g.} kinks) have compact support (so-called {\it compactons}), see \cite{Arodz02}.  A recently found non-topological object which has the form of an oscillon  is also compact, \cite{oscylony}. The other group of compact solutions which seem to be very interesting from a physical point of view are Q-balls, see \cite{Q-ball1}, \cite{Q-ball2}.  The possibility of applying Q-balls to boson stars and black holes looks especially attractive \cite{blackhole}. Also models with standard (smooth) field potential can support compact objects if they have non-standard kinetic ({\it e.g.} quartic) terms \cite{k-defects}-\cite{compsusp}. These models, known also as K-field models, are studied in the context of the global expansion of the universe (K-essence).  In fact, historically, the first non-topological compact solution was found by Roseanau and Hyman in a modified KdV system \cite{KdV}. 

The simplest V-shaped scalar field model is the signum-Gordon model \cite{scalling}, \cite{signum}. It has been shown that, apart from self-simiar solutions, it supports also so-called shock waves, {\it i.e.} a class of solutions with a field discontinuity that propagates with velocity $v=1$. The violation of the scaling symmetry of the signum-Gordon model has been studied for both self-similar solutions \cite{perturbations} and shock waves \cite{shock}. In the following paper we concentrate on the case of shock waves. The genesis of this paper comes from the observation that a specific class of solutions in \cite{shock} has not been considered. The new solutions are quite interesting because they merge properties of shock waves (discontinuity of the field at one end) and compactons (quadratic approach to a vacuum value at the other end). Such a solution is compact in an arbitrary stage of evolution so we call it {\it a shock impulse} instead of {\it a shock wave}. 

Our paper is organized as follows. In Section 2 we briefly recall the signum-Klein-Gordon model. Section 3 is devoted to the presentation of a new class of solutions. In the last Section we summarize our paper. 

\section{The signum-Klein-Gordon model}

The Lagrangian of a (1+1) dimensional signum-Klein-Gordon model reads
\begin{eqnarray}\label{lagrangian}
L=\frac{1}{2}(\partial_t\phi)^2-\frac{1}{2}(\partial_x\phi)^2 -V(\phi),
\end{eqnarray}
where $\phi(x,t)$ is a scalar field with an interaction given by the potential
\begin{eqnarray}
V(\phi)=|\phi|-\frac{1}{2}\lambda\phi^2.
\end{eqnarray} 
In \cite{shock} we have proposed a mechanical system which in the limit of an infinite number of degrees of freedom is described by the Lagrangian (\ref{lagrangian}). The quadratic interaction term has been chosen for simplicity. It is one of the simplest terms that can be used to violatate an exact scaling symetry. For details see \cite{perturbations} and \cite{shock}. The Euler-Lagrange equation reads
\begin{eqnarray}\label{EL}
\phi_{tt}-\phi_{xx}+\hbox{sign}\phi-\lambda \phi=0,
\end{eqnarray}
where we assume (for physical reasons) that sign$(0)=0$. It is clear that $\phi = 0$ is a solution of (\ref{EL}). This solution is necessary in the construction of compactons. The Ansatz $\phi(x,t)=\Theta(\pm z)W(z)$ gives discontinuous solutions, where ``$+$'' refers to solutions outside the light cone, ``$-$'' refers to solutions inside the light cone and $z=(x^2-t^2)/4$. It has been already discussed in \cite{shock} that the velocity $v=1$ is distinguished by the model and it is the unique admissible velocity for the propagation of field discontinuities. 

\section{Compact shock impulses}
We are interested in the case $\lambda \equiv \rho^2 >0$. For solutions outside the light cone our Ansatz takes the form
\begin{eqnarray}
\phi(x,t)=\Theta(z)W(z),\qquad {\rm where} \qquad z=\frac{1}{4}(x^2-t^2).
\end{eqnarray}
In the new variable $y$, related to $z$ by the formula $z=\frac{1}{4}y^2$, the equation (\ref{EL}) takes the form
\begin{eqnarray}\label{eqg}
g''+\frac{1}{y}g'+\rho^2g={\rm sign}(g),
\end{eqnarray}
where $g(y)\equiv W(z(y))$. 
It could be helpful to consider equation (\ref{eqg}) as an equation of motion of a fictious particle in the potential $U(g)=\frac{1}{2}\rho^2g^2-|g|$. As we mentioned already in  \cite{shock},  equation (\ref{eqg}) has partial solutions
\begin{equation}
g_k(y)=(-1)^k\left(\frac{1}{\rho^2}-\mu_kJ_0(\rho y)-\nu_kY_0(\rho y)\right),
\end{equation}
where $g_k>0$ for $k=0,2,4,\ldots$ and $g_k<0$ otherwise. $J_0$ and $Y_0$ are Bessel functions. The partial solution not considered in \cite{shock} is a constant zero solution. Because of sign$(0)=0$, $g(y)=0$ is a solution of (\ref{eqg}). Such a solution corresponds to the solution $\phi=0$ in a physical system, so it is well motivated from a physical point of view.

\subsection{Single-zero solution}
The solution $g_0(y)$ can be parametrized by only one parameter because at $y=0$ the Bessel function $Y_0$ has a singularity, so for physical reasons we set $\nu_0=0$. The second coefficient $\mu_0$ can be expressed by the value of $g_0(0)$, which results in 
\begin{eqnarray}
g_0(y)=\frac{1}{\rho^2}-\left(\frac{1}{\rho^2}-{g}_0(0)\right)J_0(\rho y).
\end{eqnarray} 
As was explained in a previous paper, depending on the values $g_0(0)$, the solution $g_0(y)$ can be positive or can be matched with a negative partial solution $g_1(y)$. The intermediate case allows for matching the solution $g_0(y)$ with the trivial solution $g(y)=0$. In this case $g_0(y)$ reaches its zero value quadratically. This happens for
\begin{eqnarray}
g_0(0)=\frac{1}{\rho^2}\left(1-\frac{1}{J_0(j^1_1)}\right),
\end{eqnarray}
where $j_1^1$ is the first zero of $J_1(y)$. We name this value $g_0^{crit}$. Approximately $\rho^2g_0^{crit}=3.482872$.  In this case the full  compact impulse reads
\begin{eqnarray}
g(y)= \left\{
    \begin{array}{ll}
    \frac{1}{\rho^2}\left(1-\frac{J_0(\rho y)}{J_0(j_1^1)}\right) & \textrm{if}\,\,\, 0 \leq y\leq    j_1^1/\rho\\
    0 & \textrm{if}\,\,\, y> j_1^1/\rho
    \end{array} \right.\nonumber
\end{eqnarray}

\begin{figure}[h!]
\begin{center}
\includegraphics[width=0.65\textwidth]{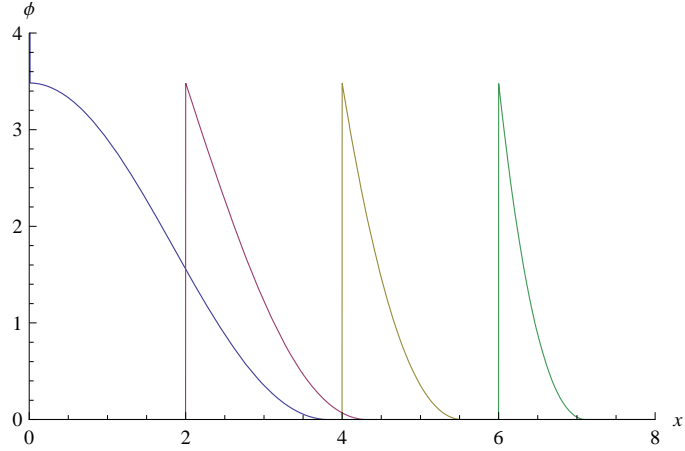}
\caption{Evolution of a compact impulse $\phi(x,t)$ at moments $t=0,2,4,6$ for $\rho=1$.}\label{evol1}
\end{center} 
\end{figure}

The matching point $x_0$ moves with subluminal velocity $\dot x_0=t/\sqrt{\overline{c}_0^2+t^2}$, where $\overline{c}_0=j_1^1/\rho$.  For $t\rightarrow \infty$ it is reached by the discontinuity $x=t$. It comes from the fact that the length of the compact shock impulse $L_c=\sqrt{\overline{c}_0^2+t^2}-t$ for $t\rightarrow \infty$ has a leading term $L_c=\frac{\overline{c}_0^2}{2}\frac{1}{t}+\mathcal{O}(\frac{1}{t^3})$.

\subsection{Multi-zero solutions}

As was discussed in \cite{shock}, for $g_0(0)>g_0^{crit}$, negative partial solutions are present. If $g_0(0)$ is not too big a full compact impulse $g(y)$ is composed of two pieces $g_0(y)$ and $g_1(y)$. Again, for a certain value of $g_0(0)$ a partial solution $g_1(y)$ reaches its zero $\overline{c}_1$ quadratically, therefore it could be matched with a constant zero partial solution. At the point $c_0$ $g_1(y)$ is matched with $g_0(y)$. We have used a special notation $\overline{c}_k$ for those zeros for which the field reaches its zero value in a quadratic way, whereas all other zeros we label by $c_k$. A general higher-rank multi-zero solution consists of partial solutions $g_0,\ldots, g_k$ and $g=0$ and has a matching points $c_0,\ldots, c_{k-1}$ and $\overline{c}_k$.

As an example let us study in details a solution for $k=1$. There are three partial solutions $g_0(y)$, $g_1(y)$ and  a constant solution $g(y)=0$, and two matching points $c_0$ and $\overline{c}_1$. The partial solutions obey the following matching conditions
\begin{eqnarray}
g_0(c_0)&=&0=g_1(c_0), \qquad g'_0(c_0)=g'_1(c_0)\label{zszycie1},\\
g_1(\overline{c}_1)&=&0,\qquad \qquad \qquad g'_1(\overline{c}_1)=0.\label{zszycie2}
\end{eqnarray}
The first condition, $g_0(c_0)=0$, and conditions (\ref{zszycie2}) give us
\begin{eqnarray}
g_0(y)&=&\frac{1}{\rho^2}\left[1-\frac{J_0(\rho y)}{J_0(\rho c_0)}\right]\label{g_0},\\
g_1(y)&=&-\frac{1}{\rho^2}\left[1-\frac{Y_1(\rho\overline{c}_1)J_0(\rho y)-J_1(\rho\overline{c}_1)Y_0(\rho y)}{Y_1(\rho\overline{c}_1)J_0(\rho \overline{c}_1)-J_1(\rho\overline{c}_1)Y_0(\rho \overline{c}_1)}\right]\label{g_1},
\end{eqnarray}
where the parameter $g_0(0)$ is expressed in terms of $c_0$. The other conditions in (\ref{zszycie1}) give algebraic equations for $c_0$ and $\overline{c}_1$. These equations can be rewritten in the following form
\begin{eqnarray}
\frac{Y_1(\rho \overline{c}_1)J_0(\rho c_0)-J_1(\rho \overline{c}_1)Y_0(\rho c_0)}{Y_1(\rho \overline{c}_1)J_0(\rho \overline{c}_1)-J_1(\rho \overline{c}_1)Y_0(\rho \overline{c}_1)}&=&1,\label{eq1}\\
\frac{Y_0(\rho c_0)}{J_0(\rho c_0)}+\frac{Y_1(\rho c_0)}{J_1(\rho c_0)}-2\frac{Y_1(\rho \overline{c}_1)}{J_1(\rho \overline{c}_1)}&=&0.\label{eq2}
\end{eqnarray} 
Equations (\ref{eq1}) and (\ref{eq2}) can be solved numerically.  
We will not present formulas for bigger  $k$ because they are too complicated.

\begin{figure}[h!]
\begin{center}
\includegraphics[width=0.6\textwidth]{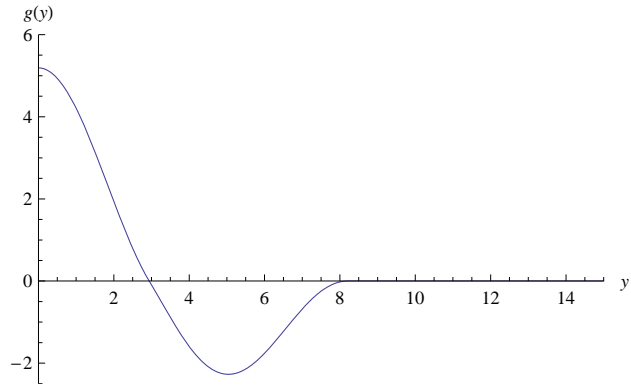}
\caption{Three segment ($k=1$) multi-zero solution $g(y)$ for $\rho=1$.}\label{multi3}
\end{center} 
\end{figure}

\begin{figure}[h!]
\begin{center}
\includegraphics[width=0.6\textwidth]{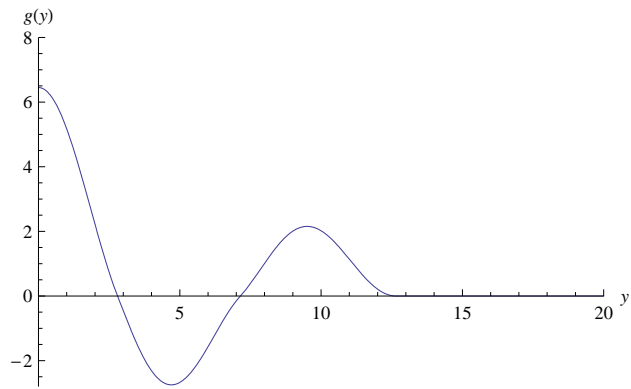}
\caption{Four segment ($k=2$) multi-zero solution $g(y)$ for $\rho=1$.}\label{multi4}
\end{center} 
\end{figure}

\begin{figure}[h!]
\begin{center}
\includegraphics[width=0.6\textwidth]{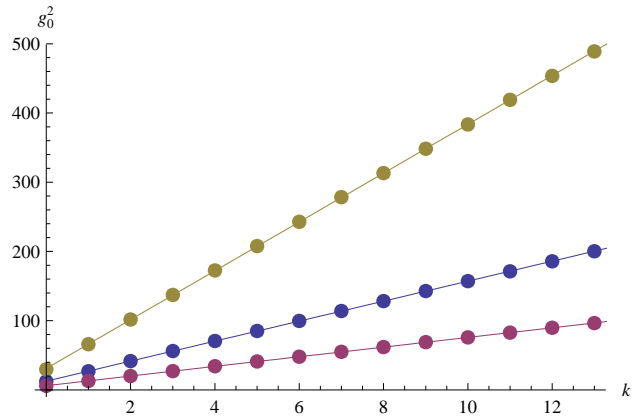}
\caption{Values of $g_0^2(0)$ for consecutive multi-zero compact solutions $g(y)$. Straight lines represent linear fits to numerical data. The upper line has been obtained for $\rho = 0.8$, the middle line for $\rho = 1.0$ and the bottom line corresponds to $\rho = 1.2$.}\label{initial_kwadrat}
\end{center} 
\end{figure}

\begin{figure}[h!]
\begin{center}
\includegraphics[width=0.6\textwidth]{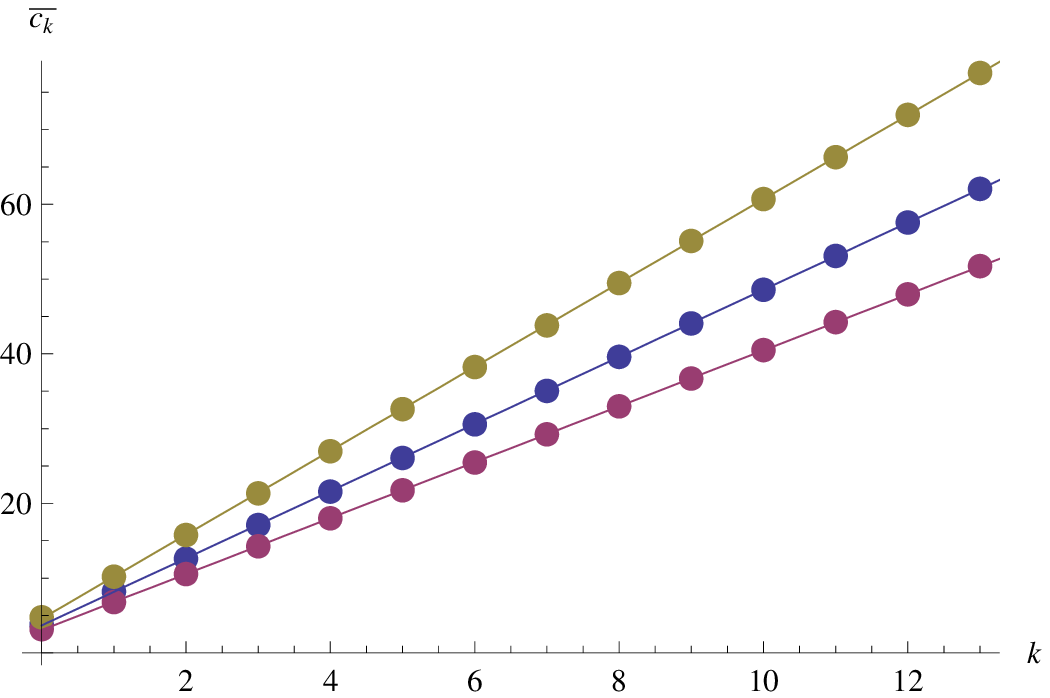}
\caption{The consecutive zeros $\overline{c}_k$ and their linear fits. The upper line has been obtained for $\rho = 0.8$, the middle line for $\rho = 1.0$ and the bottom line corresponds to $\rho = 1.2$.}\label{zeros}
\end{center} 
\end{figure}

In Figs.\ref{multi3} -\ref{multi4} we present two examples of multi-zero solutions $g(y)$ with three and four segments,  respectively. For higher-range solutions, $k>2$, instead of solutions we present only plots of numerical values of $g_0(0)$ and $\overline{c}_k$ as functions of $k$, see Figs \ref{initial_kwadrat}-\ref{zeros}. 

The numerical results suggest, surprisingly, that, at least for not too large $k$, both the sequence of $g_0^2(0)(k)$ and the sequence of zeros $\overline{c}_k$ behave linearly (arithmetic sequences).  Linear fits give us: 

\begin{itemize}
\item$\rho = 0.8$
\begin{eqnarray}
g_0^2(0)(k)&=&30.89+ 35.28 k,\nonumber\\
\overline{c}_k&=&4.57+ 5.61 k, \nonumber
\end{eqnarray}
\item$\rho = 1.0$
\begin{eqnarray}
g_0^2(0)(k)&=&12.65+ 14.44 k, \nonumber\\
\overline{c}_k&=&3.67+ 4.49 k, \nonumber
\end{eqnarray}
\item$\rho = 1.2$
\begin{eqnarray}
g_0^2(0)(k)&=&6.09+ 6.97 k, \nonumber\\
\overline{c}_k&=&3.06+ 3.74 k. \nonumber
\end{eqnarray}
\end{itemize}

This means that the size of a compact impulse (as a function of $y$) or the  size of an initial compacton configuration (as a function of $x$) changes linearly with a number $k$. The size of the compacton for a fixed $k$ shrinks when $\rho$ increases.

We want to emphasize that linear dependence $\overline{c}_k$ and $g_0(0)$ is completely unexpected. The expressions for solutions for higher $k$ and relations between coefficients are very complicated and one cannot expect such a simple relationship.

\section{Summary}
 
 We have presented a new class of solutions in the (1+1) dimensional signum-Klein-Gordon model which was not been considered in \cite{shock}. The solutions which appear in models with V-shaped potentials are usually either shock  or compacton type. Our solutions merge both of these properties. The compact shock impulses, presented above, have on one end a wave front where a scalar field is discontinuous and on the other end they approach their vacuum value quadratically.

\section{Acknowledgement}
We would like to thank H. Arodz, C. Adam, J. Sanchez-Guillen and J. Shock for discussion and valuable comments. \newline
This paper is supported by  MCyT (Spain) and FEDER (FPA2005-01963), 
and Xunta de Galicia (grant PGIDIT06PXIB296182PR and 
Conselleria de Educacion)

\end{document}